\documentstyle[11pt,newpasp,epsf,twoside]{article}
\newcommand{\HI}{\protect\normalsize H\thinspace\protect\footnotesize
I\protect\normalsize}
\newcommand{\kms}{\,km\,s$^{-1}$}

\def\edcomment#1{\iffalse\marginpar{\raggedright\sl#1\/}\else\relax\fi}
\reversemarginpar

\begin{document}
\title{Large-Scale Structures behind the Southern Milky Way in the Great 
Attractor Region}
\author{Patrick A. Woudt}
\affil{Department of Astronomy, University of Cape Town, Private Bag, 
Rondebosch 7700, South Africa}
\author{Ren\'ee C. Kraan-Korteweg}
\affil{Depto.\ de Astronom\' \i a, Universidad de Guanajuato, 
Apdo.~Postal 144, Guanajuato GTO 36000, Mexico}

\begin{abstract}
A deep optical galaxy search behind the southern Milky Way and a
subsequent redshift survey of the identified obscured galaxies traces
clusters and superclusters into the deepest layers of Galactic
foreground extinction (${A_B} \le 3^{\rm m} - 5^{\rm m}$).  In the
Great Attractor region, we have identified a low-mass cluster (the
Centaurus--Crux cluster) at (${\ell}, b, v, {\sigma}$) = ($305\fdg5,
+5\fdg5, 6214$\,\kms, $472$\,\kms) and found that ACO 3627 (the Norma
cluster) at (${\ell}, b, v, {\sigma}$) = ($325\fdg3\, -7\fdg2,
4844$\,\kms, $848$\,\kms) is the most massive cluster in the Great
Attractor region known to date.  It is comparable in virial mass,
richness and size to the well-known but more distant Coma cluster. The
Norma cluster most likely marks the bottom of the potential well of
the Great Attractor.  It is located at the intersection of two main
large-scale structures, the Centaurus Wall and the Norma
supercluster. The flow field observed around the Great Attractor probably
is caused by the confluence of these two massive structures.
\end{abstract}

\section{The Zone of Avoidance and the Great Attractor}

The Milky Way acts as a natural barrier for studies of the large-scale
structure and large-scale dynamics in the Universe; dust and stars in
the disk of the Milky Way obscure the light of $\sim$20\% of the
optical extragalactic sky. Galaxies appear smaller and fainter towards
the plane of the Milky Way due to the increasing Galactic extinction.
This renders diameter- and magnitude-limited samples of galaxies highly
incomplete near the Galactic Plane, resulting in a `Zone of
Avoidance' (ZOA) in the distribution of galaxies.

The Great Attractor (GA) -- a dynamically important overdensity of galaxies
in the local Universe -- is located close to, or behind, the southern
Milky Way (Lynden-Bell et al.~1988; Kolatt, Dekel, \& Lahav 1995; 
Tonry et al.~2000).

This nearby mass overdensity was first noticed through a large-scale
systematic flow of galaxies (Lynden-Bell et al.~1988). Even though
there is a large excess of optical and IRAS galaxies in this region
(Lynden-Bell 1991), no dominant cluster or central peak had been
identified (but see Kraan-Korteweg et al.~1996; Woudt 1998), which
strongly suggests that a significant fraction of the Great Attractor
overdensity could still be obscured by our Milky Way.

\section{The Galaxy Distribution in the Great Attractor Region}

There are various approaches to trace the galaxy distribution behind
the southern Milky Way. One can infer the galaxy distribution
indirectly by analysing the peculiar motions of galaxies, assuming
that these arise from the irregular mass density field in the local
Universe.  Alternatively, one can try to {\sl observe} the galaxy
distribution at various wavelengths.  By directly observing the galaxy
distribution in the ZOA and comparing this with the reconstructed mass
density field, important cosmological parameters such as $\Omega_0$
can be inferred (Dekel 1994; Sigad et al.~1998).

\subsubsection{The Surmised Galaxy Distribution in the GA Region.} 

A comprehensive review of the various reconstruction methods, their
results, and a comparison with the observed distribution of unveiled
galaxies in the ZOA is given by Kraan-Korteweg \& Lahav (2000). The
two most commonly used methods to reconstruct the galaxy distribution
behind the Milky Way are the Wiener Filter method, a statistical
approach (see Hoffman; Zaroubi, these proceedings), and the POTENT
method (Bertschinger \& Dekel 1989; Dekel 1994). The latter uses the
peculiar velocities of galaxies to reconstruct the mass density
field. The Great Attractor is clearly visible as a significant mass
excess located close to the southern Milky Way in both reconstructions
(Bistolas 1998; Webster, Lahav, \& Fisher 1997; Kolatt et al.~1995).
In fact, the best evidence for the existence of the Great Attractor
comes from the reconstructed galaxy/mass distribution.  But the less
than perfect match between the reconstructed mass density field and
the observed galaxy distribution (Dekel 1994) has led to contradictory
interpretations of the true nature and extent of the Great Attractor
(Dressler 1988; Hudson 1993a, 1993b; Jahoda \& Mushotzky 1989;
Rowan-Robinson et al.~1990; Lynden-Bell 1991).

\subsubsection{The Observed Galaxy Distribution in the GA Region.}

Dedicated deep multiwavelength surveys -- optical, near infrared
(NIR), far infrared (FIR), \HI, X-ray -- were initiated, to search for
a galaxy excess associated with the Great Attractor. For detailed
information about galaxy searches in the GA region at wavelengths
other than the optical, see the contributions by Schr\"oder et
al.~(NIR), Saunders et al.~(FIR), Staveley-Smith et al.~(\HI),
B\"ohringer et al.~(X-ray) and Ebeling et al.~(X-ray) in these
proceedings.  Many of these searches are complementary in the galaxy
population they trace.  An overview of the advantages and limitations
of each of these methods is given by Woudt (1998).

\subsection{A Deep Optical Galaxy Search in the GA Region.}

For many years, we have been involved in a deep optical galaxy search
behind the southern Milky Way (Kraan-Korteweg 1989; Kraan-Korteweg
2000; Fairall, these proceedings). Here, we report on results
in the Crux and Great Attractor region (Kraan-Korteweg \& Woudt 1994;
Woudt 1998).  This galaxy survey was made on the IIIaJ film copies of
the SRC Sky Survey. It has a diameter limit of ${D} \ge 0\farcm2$. In
the Crux and Great Attractor region ($295\deg \le \ell \le 340\deg$
and $-10\deg \le b \le +10\deg$), 8182 galaxies have been identified
(Woudt 1998). Only 2.3\% were catalogued before (Lauberts
1982).  Our ZOA survey is complete for all galaxies with an
extinction-corrected diameter ${D^0} \ge 1\farcm3$, down to extinction
levels of ${A_B} \le 3^{\rm m}$ (Woudt 1998; Kraan-Korteweg 2000,
Kraan-Korteweg, these proceedings).

\begin{figure}
\plotfiddle{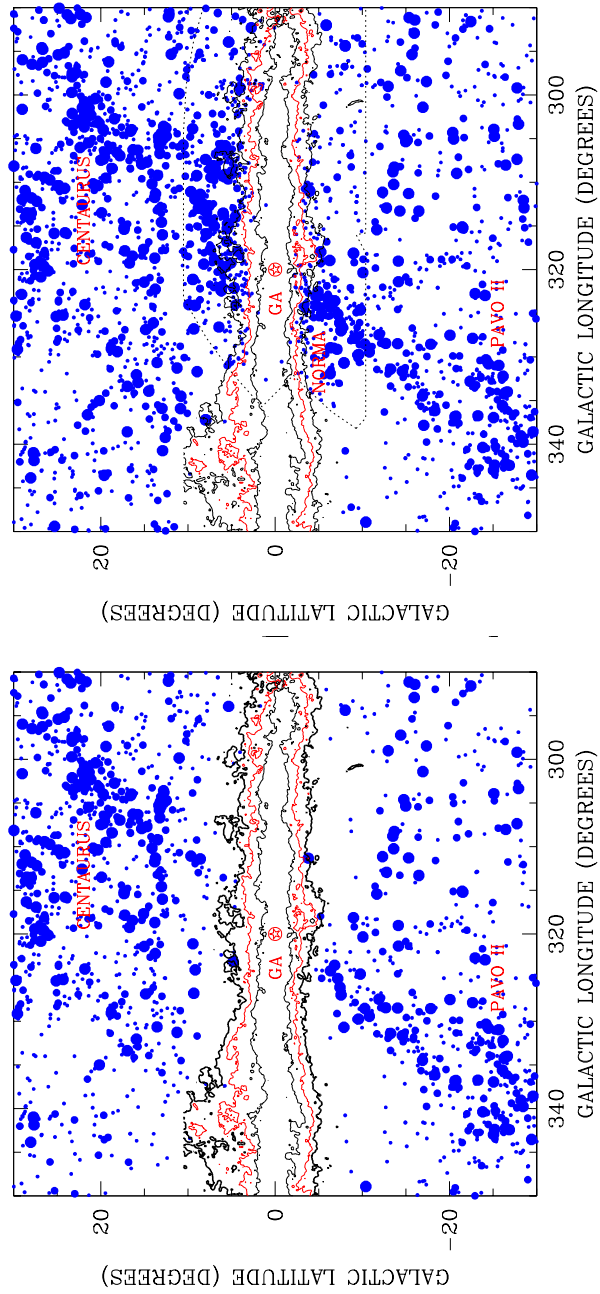}{16.4cm}{0}{64}{64}{-200}{-10}
\caption{The distribution of galaxies with observed major diameters
${D} \ge 1\farcm3$ (left panel) and {\sl extinction-corrected} major
diameters ${D^0} \ge 1\farcm3$ (right panel) in the general Great
Attractor region. The right panel includes the data from our deep
optical galaxy survey (delineated area). The contours mark extinctions
of ${A_B} = 3.0^{\rm m}$ (thick line), $5.0^{\rm m}$ and $10.0^{\rm
m}$ (Schlegel et al.~1998).  The galaxies are diameter-coded, the
galaxies with $1\farcm3 \le {D} \le 2'$ are displayed as points, the
galaxies with $2' \le {D} \le 3'$ as small circles and the galaxies
with ${D} \ge 3'$ are shown as large circles.}
\end{figure}

The left panel in Fig.~1 shows the distribution of Lauberts (1982)
galaxies in the general direction of the Great Attractor. The galaxies
are diameter-coded. Only galaxies with an observed diameter larger
than $1\farcm3$ -- the completeness limit of the Lauberts catalogue
(Hudson and Lynden-Bell 1991) -- are included. The influence of the
Galactic foreground extinction is obvious, the Lauberts catalogue is
at best complete down to $|b| \approx 10\deg$ though probably
incomplete already between $|b| = 10\deg - 15\deg$ (where ${A_B} \ge
1\fm0$) for galaxies with ${D} \ge 1\farcm3$.

In the right panel of Fig.~1, all the Laubert galaxies have been
corrected -- following Cameron (1990) -- for the obscuring effects of
the Galactic extinction using the DIRBE/IRAS extinctions (Schlegel,
Finkbeiner, \& Davis 1998). Only the galaxies with an {\sl
extinction-corrected} diameter (${D^0}$) larger than $1\farcm3$ are
plotted.  This diagram also includes the galaxies from our deep
optical galaxy search which have ${D^0} \ge 1\farcm3$. The galaxies
below our completeness limit (at ${A_B} \ge 3^{\rm m}$, the outer
thick contour in Fig.~1) are shown, but they are not diameter-corrected. 
The diameter corrections
(Cameron 1990) become somewhat uncertain at those extinction levels.
The right panel of Fig.~1 shows a significant excess of galaxies north
of the Galactic Plane, at $305\deg \le \ell \le 323\deg$ and $b =
4\deg - 10\deg$, that went unnoticed before (left panel). The highest
concentration of galaxies in our entire deep optical galaxy search is
centred on ACO 3627 (Abell, Corwin, \& Olowin 1989), the Norma
cluster, at $(\ell, b)$ = $(325\fdg3, -7\fdg2)$. This cluster lies
within $10\deg$ of the predicted centre of the Great Attractor (Kolatt
et al.~1995). The richness of this centrally condensed cluster had not
been remarked because of the diminishing effects of the Galactic
foreground extinction.

Our ZOA survey shows that the Great Attractor region is overdense in
galaxies down to very low Galactic latitude, with further prospective
(super-) cluster members still hidden behind the remaining optical ZOA
of ${A_B} \ge 3^{\rm m}$ (but see Staveley-Smith et al., these proceedings).

\section{The Results of our ZOA Redshift Survey}

A redshift coverage of the nearby galaxy population in the ZOA is
essential for the determination of the peculiar velocity of the Local
Group and for the mapping of the velocity flow field. With our
redshift survey we have aimed to obtain a fairly homogeneous and
complete coverage of the brighter and larger galaxies found by our
deep optical survey. Three distinctly different observational
approaches were used: {\bf 1.}~optical spectroscopy for individual
early-type galaxies (Kraan-Korteweg, Fairall, \& Balkowski 1995;
Fairall, Woudt, \& Kraan-Korteweg 1998; Woudt, Kraan-Korteweg, \&
Fairall 1999), {\bf 2.}~low resolution, multi-fibre spectroscopy for
the high-density regions (see Felenbok et al.~1997, Woudt~1998), and
{\bf 3.}~\HI\ line spectroscopy for spiral galaxies (Kraan-Korteweg,
Woudt, \& Henning 1997).

For 1211 galaxies -- about 15\% of the optically detected galaxies in
the Crux and Great Attractor region -- a reliable redshift has now been
recorded. In this part of the sky, 134 redshifts were known
previously. Voids, clusters and superclusters can now be clearly
identified in the Zone of Avoidance.

Figure 2 shows the distribution of galaxies in redshift-space (0 -- 8000~\kms) 
centred on the Great Attractor.  In the first slice (0 --
2000\,\kms), the Supergalactic Plane is running nearly vertically
across the diagram from $\ell = 315\deg$, $b = 30\deg$ to $\ell =
335\deg$, $b = -30\deg$.  In the second slice ($2000 \le v \le
4000$\,\kms), the new data reinforce the presence of a narrow
filamentary structure running all the way from $\ell = 340\deg$, $b =
-25\deg$ to the Centaurus cluster at $\ell = 303\deg$, $b =
20\deg$. This is believed to be part of a Great Wall-like structure
seen edge-on, the Centaurus Wall (Fairall et al.~1998). The new
observations fill in a significant part of this feature. Next to the
Centaurus Wall, a broad band of galaxies runs nearly vertically across
the GP (at $\ell = 310\deg$) towards an overdensity of galaxies at
$\ell \approx 310\deg$, $b \approx -15\deg$. This extended filament is
located between the Centaurus Wall and the Hydra--Antlia
supercluster. It is not clear at this stage, if this overdensity is
part of the Centaurus Wall.  At the location of ACO 3627 (the Norma
cluster), a concentration of galaxies is evident, but this is due to
the low velocity tail of the cluster with its large velocity
dispersion (848\,\kms; Woudt 1998).

The Norma cluster at (${\ell}, b, v$) = ($325\fdg3, -7\fdg2,
4844$\,\kms) becomes very pronounced in the third redshift slice
($4000 \le v \le 6000$\,\kms), where it is the dominant feature in
this part of the sky. This redshift slice corresponds to the
redshift-distance of the Great Attractor overdensity, i.e.~$\sim
4500$\,\kms\ (Lynden-Bell et al.~1988; Kolatt et al.~1995).  This
slice also corresponds to the strong single peak seen in the velocity
histogram of the GA region and of part of the broad peak seen in the
Crux region (Woudt 1998).  The new data -- with the galaxies around the
Pavo II cluster ($\ell \approx 332\deg$, $b \approx -23\deg$) adjacent
to the survey -- suggest a large-scale structure that runs more or less
horizontally across the diagram (Woudt 1998).  This new broad feature
will hereafter be referred to as the ``Norma supercluster''.  Traces
can also be seen in the following slice, so the structure is probably
also wall-like, seen roughly side-on. Its width (or depth in Fig.~2)
is some 3000\,\kms\ and its thickness several hunderd \kms. Bearing in
mind its greater distance compared to the Centaurus Wall mentioned
above, this new structure must be similarly massive. It also covers a
smaller cluster/group of galaxies around $\ell = 305\deg, b = +5\deg$
at $v = 6214$\,\kms\ (Fairall et al.~1998). Both this Centaurus--Crux
cluster, and the Vela overdensity ($\ell = 280\deg, b = 6\deg, v =
6000$\,\kms; Kraan-Korteweg \& Woudt 1993) probably form part of the
Norma supercluster.  The Norma cluster itself is situated where these
two massive structures -- the Centaurus Wall and the Norma
supercluster -- intersect.

One cannot exclude the possibility that another (Coma-like) rich
galaxy cluster is still hidden in the GA region behind the remaining
optical ZOA. At these higher extinction levels galaxies are optically
invisible. But the central galaxy of a rich cluster is likely to be a
strong radio continuum source, hence observable.  Exactly such an
object, PKS\,1343$-$601, lies in the GA region at (${\ell}, b, v,
{A_B}$) = ($309\fdg7, +1\fdg8, 3800$\,\kms, 12$^{\rm m}$; West \&
Tarengi 1989; Woudt 1998; Kraan-Korteweg \& Woudt 1999).  There are
various indications that this radio galaxy indeed marks the position
of a rich cluster (see Kraan-Korteweg \& Woudt 1999). Finding another
rich cluster in the GA region would have serious implications for our
understanding of this massive overdensity.

\vspace{1cm}
\begin{figure}[h]
\plotfiddle{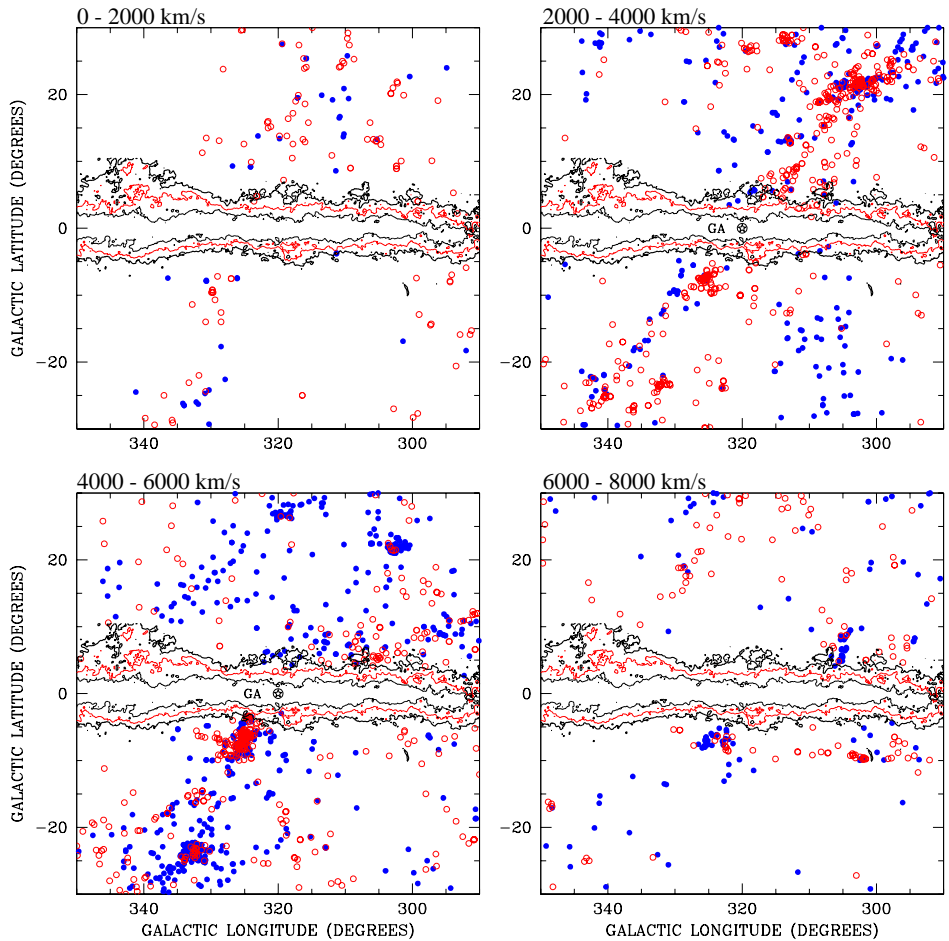}{11.70cm}{0}{65}{65}{-195}{-80}
\caption{Sky projections in Galactic coordinates for redshift interval
of $\Delta v = 2000$\,\kms\ centred on the Great Attractor. Within the
panels the redshifts are subdivided into intervals of $\Delta v =
1000$\,\kms: the filled dots mark the nearer redshift interval (e.g.,
$v \le 1000$\,\kms\ in the top-left panel), the open circles the more
distant interval ($1000 \le v \le 2000$\,\kms\ in the same panel). The
skyplots increase in velocity-distance from the top-left panel to the
bottom-right panel as marked above each panel. The contour levels of
the Galactic foreground extinction are as in Fig.~1.}
\end{figure}

\section{Measuring Peculiar Motions of Galaxies in the ZOA}

The Norma cluster can be used to address some of the remaining
outstanding issues with regard to the nature and extent of the Great
Attractor. Is the cluster located at the bottom of the potential well
of the Great Attractor?  Is the Great Attractor a relatively nearby
overdensity at $\sim$3000\,\kms\ as suggested by Tonry et al.~(2000),
or is it more distant at $\sim$4500\,\kms\ (e.g.~Kolatt et
al.~1995). Does the Great Attractor itself partake in a flow on even
larger scales?

In order to address these questions, peculiar motions of galaxies {\sl
behind} the Milky Way are needed.  Reliable peculiar motions of ZOA
galaxies are not easy to obtain, the problems are severe and need
careful consideration (McCall \& Armour, these proceedings).
Uncertainties in the Galactic foreground extinction (Hudson 1999), the
correction of observed parameters (isophotal diameters, magnitudes)
(Cameron 1990), the effect of star-crowding on the photometry of
partially obscured galaxies (Buta \& McCall 1999) could result in
large systematic uncertainties in the measured peculiar velocity of
ZOA galaxies.

\subsubsection{The Galactic Foreground Extinction.}

The DIRBE/IRAS reddening maps provide an accurate measure of the
Galactic foreground extinction (Schlegel et al.~1998), and are a
significant improvement over the reddening estimates derived from the
Galactic neutral hydrogen column density maps, especially in the Great
Attractor region where large variations in the \HI\ gas-to-dust ratio
are suspected (Burstein et al.~1987).  However, the accuracy of the
DIRBE/IRAS maps at very low Galactic latitudes ($|b| \le 10\deg$)
still has to be established, as they are only calibrated by reddening
measurements away from the Galactic Plane (Schlegel et al.~1998). A
pilot study by Woudt (1998), using 18 elliptical/lenticular galaxies
in the ZOA and the colour-Mg$_2$ relation (Bender, Burstein, \& Faber
1993), has shown that the DIRBE/IRAS maps provide a good estimate of
the Galactic extinction in the ZOA, at least down to ${E(B-V)} \le 0.5$
mag.  This pilot study is currently extended to cover the entire
southern Milky Way and a larger range of Galactic reddening values. It
will allow a calibration for the DIRBE/IRAS maps at low Galactic
latitudes.

\subsubsection{The Peculiar Motion of ACO 3627.}

Given its nature as a rich cluster and its unique location at the
centre of the GA, the Norma cluster is the prime target to start and
explore these uncertainties.  The first attempt in deriving a peculiar
velocity of the Norma cluster (Mould et al.~1991) has been unsuccesful
but illustrates the difficulties involved. An unrealistically large
peculiar motion of 1760 $\pm$ 355\,\kms\ was derived from 7 galaxies
in the Norma cluster, using the $I$-band Tully-Fisher
relation. Following a detailed dynamical analysis of the Norma cluster,
Woudt (1998) has shown that the former result was strongly biased
because galaxies from an infalling spiral-rich subgroup were included
(3 of the 7 selected galaxies). Even if the Galactic foreground
extinction is accurately known, the complexity of the Norma cluster
does not allow an accurate distance determination based on 7 galaxies
alone.

By using a sample of 50 elliptical and lenticular galaxies within the
inner part of the Norma cluster and the $R$ and $K'$ Fundamental Plane
analysis (Pahre, Djorgovski, \& de Carvalho 1998; Mobasher et al.~1999)
an accurate distance measurement ($\sim$3\% accuracy, or 150\,\kms\ at
the distance of the Norma cluster) can be obtained.  Moreover, the $(R
- K')$ colours of these galaxies, combined with their Mg$_2$ indices,
will provide an accurate measure of the extinction (Bender et
al.~1993) independent of the assumed reddening law. High
signal-to-noise spectra have already been obtained, as well as deep
$R$-band photometry of the entire Norma cluster within its Abell
radius. $K'$-band photometry will be obtained with the NTT during 4
nights in June 2000.  These data will allow us to distinguish in a
quantitative way between the various existing models of the Great
Attractor.

\section{The Great Attractor Unveiled?}

Our view of the Great Attractor has been severely obscured by the
Milky Way.  The here reported deep optical galaxy search has revealed
a large fraction of the GA that was previously hidden by the
Galaxy. This was done by the simple means of eye-balling the IIIaJ
film copies. An optical ZOA remains (${A_B} \ge 3^{\rm m} - 5^{\rm
m}$). There studies in the near-infrared and at 21-cm can trace
the GA across the deepest layers of the Galactic extinction.

The emerging optical picture of the Great Attractor is that of a
confluence of superclusters (the Centaurus Wall and the Norma
supercluster) with the Norma cluster being the most likely candidate
for the Great Attractor's previously unseen centre. However, the
potential well of the GA might be rather shallow and extended, and
another (highly obscured) rich cluster around the strong radio-source
PKS1343$-$601 could also be located at the bottom of the GA's
potential well.

\bigskip
\noindent
\acknowledgements
We kindly acknowledge Drs. A.P. Fairall, C. Balkowski, V. Cayatte and P.A. Henning
for their collaborational efforts in the ZOA redshift survey.

\end{document}